\documentstyle[aps,preprint,tighten]{revtex}

\newcommand{\bx}{{\bf x}}
\newcommand{\br}{{\bf r}}
\newcommand{\bk}{{\bf k}}

\newcommand{\bnab}{\mbox{\boldmath $\nabla$}}
\newcommand{\fith}{\widehat\varphi}
\newcommand{\gtil}{\widetilde{G}}
\newcommand{\gatil}{\widetilde{\Gamma}}
\newcommand{\etil}{\widetilde{\eta}}
\newcommand{\bkappan}{\mbox{\boldmath $\kappa$}_n}
\newcommand{\bkappa}{\mbox{\boldmath $\kappa$}}
\newcommand{\barx}{\bar{x}}

\draft

\begin{document}

\title{\bf Thermal and Quantum Fluctuations around Domain Walls}
\author{Carlos A. A. de Carvalho\footnote{aragao@if.ufrj.br}}
\address{Instituto de F\'\i sica, Universidade Federal do Rio de Janeiro\\
C.P.~68528, Rio de Janeiro, RJ 21941-972, Brasil}
\date{\today}
\maketitle


\begin{abstract}

We compute thermal and quantum fluctuations in the background of a domain wall
in a scalar field theory at finite temperature using the {\em exact} scalar propagator 
in the subspace orthogonal to the wall's translational mode. The propagator makes 
it possible to calculate terms of any order in the semiclassical expansion of the 
partition function of the system. The leading term in the expansion corresponds to 
the fluctuation determinant, which we compute for {\em arbitrary} temperature in space 
dimensions $1$, $2$, and $3$. Our results may be applied to the description of thermal scalar 
propagation in the presence of soliton defects (in polymers, magnetic 
materials, etc.) and interfaces which are characterized by kinklike profiles.
They lead to predictions as to how classical free energy differences, surface tensions,
and interface profiles are modified by fluctuations, allowing for comparison
with both numerical and experimental data. They can also be used to estimate
transition temperatures. Furthermore, the simple analytic form of the propagator may simplify 
existing calculations, and allow for more direct comparisons with data from scattering experiments.

\end{abstract}



\section{Introduction}
\label{introduction}

There are many problems in physics which require quantization around
classical backgrounds, such as Schwinger's QED calculation of vacuum
polarization around constant electromagnetic fields \cite{schwinger},
the semiclassical quantization of field theories \cite{dhn}, and the
semiclassical analysis of the {\em classical} limit
\cite{gutzwiller,berry} of quantum mechanics. In fact, the latter 
two characterize the semiclassical analysis
as a two-way road: towards quantization, for the field theorists;
towards the classical limit, for  learning how classical
physics descends from the quantum world. 

Taking the field theory lane of that two-way road, we have recently
studied the problem of quantizing the simplest of field theories,
one-dimensional quantum mechanics, whose (scalar) ``field''
lives in zero spatial dimension \cite{semi,bsas,pre1,pre2}. Working in
{\em imaginary time}, i.e., at finite temperature, and with smooth
potentials, we were able to show that a full semiclassical series
could be constructed from the knowledge of semiclassical propagators 
in the classical backgrounds of the solutions of the equations of motion, 
and gave a general prescription for deriving them from those
backgrounds. Our construction was similar to the one proposed long ago
by DeWitt-Morette \cite{dwm} in {\em real time}, i.e., at zero temperature: 
it generalized to quantum statistical mechanics, where the number of
classical solutions is drastically reduced, the results for ordinary
quantum mechanics. 

This article applies some of our techniques to semiclassical quantization
around a very specific background in a very specific scalar field theory, at
finite temperature, in one, two, and three spatial dimensions. In that
respect, it bears greater similarity to the Schwinger problem than
to the semiclassical quantization of field theories. The latter 
would require finding all classical solutions that satisfy the finite 
temperature boundary conditions, and a general procedure for constructing 
semiclassical propagators in their backgrounds. As we have already stated,
in quantum mechanics the general procedure does exist, and it
is often possible to find all classical solutions \cite{semi,bsas,pre1,pre2}.
In field theories, however, finding all classical solutions is already
a difficult problem, and even if we find them all, no general procedure to obtain 
semiclassical propagators in their backgrounds is known. The technical reason for 
that is the presence of spatial dimensions, which transform the fluctuation problem, 
given by an ordinary differential equation (ODE) in zero spatial dimension, 
into a partial differential equation (PDE) problem, whose solutions cannot 
be obtained from the mere knowledge of the background. Even finding the 
backgrounds involves, in general, a PDE, instead of the ODE of quantum mechanics. 

We have thus examined a specific potential of the $\varphi^4$-type, and confined 
our analysis to a (Euclidean) time-independent solution of the classical
equations of motion depending on only one spatial coordinate, a domain
wall whose classical profile and dynamics of fluctuations could be
{\em exactly} solved from ODE's. We were then able to construct a semiclassical 
propagator in the presence of that background, from which,
in principle, we can generate any term of the semiclassical series
around the domain wall. We note that even in one spatial dimension, where the 
domain wall has finite Euclidean action, and therefore contributes to 
the semiclassical approximation, we abdicate the idea of ``solving'' the 
theory semiclassically, because there are many other solutions whose 
semiclassical propagators cannot be obtained from a general recipe. In
particular, solutions which only depend on Euclidean time do contribute:
they are given by elliptic functions, similar to the ones in quantum mechanics
\cite{semi,bsas,pre1,pre2}, but whose fluctuation problems are much harder
to solve.

Our motivation for pursuing this problem, besides the technical appeal 
of obtaining thermal and quantum fluctuations at any temperature
from an {\em exact} expression for the semiclassical propagator, was  
dictated by its potential physical applications. Indeed, for spatial dimension
$d=1$, our formalism can be adapted to describe fluctuations around
defects that occur in one-dimensional systems, such as the {\em
solitons} found in polymers (like polyacetylene) \cite{polymers}; and for
$d=2,3$, to describe fluctuations around {\em interfaces} that
separate two distinct phases \cite{interfaces}, a description that allows one
to calculate how the surface tension, as well as the interface 
profile are modified by those fluctuations. It can also be used to approximately 
estimate transition temperatures \cite{dwft}. 
Adopting this view, we are able to extract physically relevant results from the study 
of $d>1$, where the action of the domain wall diverges with the size of the system. 

Since the problem is treated at finite temperature $T$, fluctuations
incorporate thermal and quantum effects, which justifies the term
thermal and quantum fluctuations employed thus far. The background we use,
however, does not depend on $T$, as it is a ``static''
classical solution of the equations of motion, independent of the
Euclidean time. That independence guarantees that it remains a {\em classical}
solution at {\em any} temperature $T$, which is all we require to compute
a semiclassical expansion order by order. As we incorporate thermal effects
through fluctuations, our classical solution will no longer be a minimum of
the (now $T$-dependent) Gibbs free energy (the effective action). Nevertheless,
the new minimum may be obtained by extremizing the free energy with an ansatz
that adds corrections to the original kink profile. The procedure will be outlined in
Section \ref{physappl}, and will profit from the nice properties of that original profile. Physically, in order for the procedure to be justified,
we must be dealing with situations where we start with a rather ``massive'' 
background, but whose fluctuations are ``light'', when both are compared to the
temperature $T$. Thus, only the fluctuations would be influenced by the interaction with the heat bath, as in our calculations. This is very much what would be expected
in the context of the so-called ``adiabatic'' approximations, where
one can distinguish ``heavier'' and ``lighter'' degrees of freedom, as
in the physics of polymers. Clearly, we have to assume that boundary conditons
at spatial infinity of the domain wall type are externally fixed, which forces us
to go to the kink sector.

We should stress that the problem of computing fluctuations around
domain wall solutions has played an important role in the physics of
interfaces in binary mixtures \cite{interfaces}. In that case,
however, temperature was introduced in a phenomenological way, through
the ``coarse-grained'' Ginzburg-Landau free energy, whose mass parameter was
assumed temperature dependent. In order to avoid confusion, we will call 
that temperature $T_{\rm ph}$. Our problem will reduce to that one if
we also assume that the mass parameter of our Lagrangian depends on
$T_{\rm ph}$, in a manner determined phenomenologically, and, furthermore, 
that our fluctuations are coupled to a heat bath at {\em zero} temperature,
$T=0$. Then, our methods can be used in the calculations
that compute critical exponents to two-loop order \cite{munster},
which obtain excellent agreement with both numerical and real experiments.
Those calculations make use of the semiclassical propagator in the form
of a series over eigenmodes. Our closed expression should simplify them,
and should allow for an independent check. We, however, treat a different physical
situation: one in which a microscopic field theory is forced to have two
regions at distinct vacua. For that case, we compute how the interface
separating those regions is affected when a heat bath at $T$ couples to its quantum 
fluctuations. In the polymer context, at $d=1$, that amounts to deriving how 
thermal phonons modify a soliton defect. For $d>1$, that can tell us how
thermal neutral bosons affect an interface separating regions in 
different Bose-condensed phases.
  
We should point out that a number of works have dealt with similar questions:
the finite temperature determinant has been computed long ago \cite{dwft,bg},
as a product over eigenvalues;
Parnachev and Yaffe \cite{larry} have carefully studied the same
problem at $T=0$, paying special attention to the determinant
calculation using a different method.
Graham {\em et al.\/} \cite{graham} have investigated similar
problems involving interfaces, resorting to the relation with
phase shifts, rather than with the semiclassical propagator.
In all cases, it seems possible to establish a precise
relation to the approach we adopt. It should be noted that
our approach includes a finite temperature $T$,
and derives the {\em exact} semiclassical propagator which,
together with a careful treament of collective coordinates,
should allow for the computation of correlation functions
beyond leading order.

The structure of the article is as follows: in
Section \ref{sidw}, we briefly review how one derives the
semiclassical expansion; Section \ref{propagator} obtains
an {\em exact} analytic expression for the semiclassical
propagator, while Section \ref{determinants} uses it to
compute fluctuation determinants; in
Section \ref{renormalization}, we describe the renormalization
procedure, first by regularizing the calculation in
Section \ref{regularization}, and then by removing its
divergences in Section \ref{renorm}; in Section \ref{collective},
we discuss the collective coordinates associated with the
translational invariance of the solution; in Section \ref{physappl},
we show how surface tensions and interface profiles are modified by
fluctuations, and outline how the first higher order correction can be computed.
We also mention how transition temperatures can be estimated from our results. 
We present our conclusions in Section \ref{conclusions}.


\section{The semiclassical expansion}
\label{sidw}

The partition function for a self-interacting scalar field theory
in contact with a thermal reservoir at temperature $T$ can be written as 
($\beta=1/T$)
\begin{equation}
\label{Z1}
Z(\beta)=\oint
[{\cal D}\varphi]\, e^{-S[\varphi]/\lambda}\;,
\end{equation}
\begin{equation}
S[\varphi]=\int_0^{\beta} d\tau \int d^d x\, {\cal L}[\varphi]\;,
\end{equation}
\begin{equation}
\label{LAGRANGEAN}
{\cal L}[\varphi]=\frac{1}{2}(\partial_{\tau} \varphi)^2
+ \frac{1}{2}(\bnab \varphi)^2 +
\frac{1}{4!}(\varphi^2-\varphi_v^2)^2\;.
\end{equation}
The integral $\oint$ is to be performed over all $\varphi$'s
that satisfy the boundary conditions
$\varphi(0,\bx)=\varphi(\beta,\bx)$. The fields are rescaled
so that the coupling constant $\lambda$ appears as an overall factor.

We shall be interested in domain wall ans\"atze $\fith(x)$ which
only depend on one spatial coordinate $x\equiv x^1$, which will be called 
longitudinal. As is well known \cite{dhn}, the equation of motion from 
(\ref{LAGRANGEAN}) is satisfied by a kinklike profile:
\begin{equation}
\fith(x)=\pm \varphi_v \tanh\left[\frac{\varphi_v (x - \bar{x})}
{2\sqrt{3}}\right],
\end{equation}
where $\bar{x}$ denotes the position of the domain wall
separating two spatial regions of dimension $d$. The remaining
$d-1$ transverse spatial coordinates of $\bx$ define a vector $\br$. The
$\pm$ refers to kink and antikink, respectively. The classical
action of the wall is proportional to the volume of the
$(d-1)$-dimensional subspace orthogonal to the $x$-axis, which we will 
denote by $V_{d-1}$ ($V_{0}\equiv 1$):
\begin{equation}
S[\fith]=\frac{2}{3\sqrt{3}}\beta\, \varphi_v^3\, V_{d-1}\;.
\end{equation}

We expand the action around the domain wall configuration $\fith$
in order to derive the semiclassical series in that background:
\begin{equation}
\label{expansion}
\varphi(\tau,\bx)=\fith(\tau,\bx) + \lambda^{1/2}\eta(\tau,\bx)\;,
\end{equation}
\begin{equation}
\label{BCETA}
\eta(0,\bx)=\eta(\beta,\bx)=0\;.
\end{equation}
As usual, the fluctuation $\eta$ has to vanish at $\tau=0$
and $\tau=\beta$ because $\fith$ already satisfies the boundary
condition. The expression for the partition function becomes
\begin{equation}
\label{Z2}
Z(\beta)= e^{-S[\fith]/\lambda}\oint [{\cal D}\eta] e^{-(S_2+\delta
S)}\;,
\end{equation}
where both $S_2$ and $\delta S$ depend functionally on $\fith$
and on $\eta$:
\begin{equation}
\label{S2}
S_2[\fith,\eta]=\int_0^{\beta} d\tau \int d^d x\,
\left\{\frac{1}{2}(\partial_{\tau} \eta)^2 + \frac{1}{2}(\bnab \eta)^2
+ \frac{1}{12}(3\fith^2-\varphi_v^2)\,\eta^2\right\}\;,
\end{equation}
\begin{equation}
\delta S[\fith,\eta]=\int_0^{\beta} d\tau \int d^d x\,
\left\{ \frac{\lambda^{1/2}}{3!}\fith\,\eta^3
+\frac{\lambda}{4!}\eta^4\right\}\;.
\end{equation}
If we define $M^2\equiv \varphi_v^2/3$, and use the explicit
form of $\fith$, the quadratic Lagrangian in (\ref{S2}) can be written
\begin{equation}
\label{L2}
{\cal L}_2=\frac{1}{2}\left\{(\partial_{\tau} \eta)^2
+ (\bnab \eta)^2 + [M^2 - \frac{3}{2}M^2{\rm sech}^{2}(\xi)]\,
\eta^2\right\}\;,
\end{equation}
with $\xi\equiv [\varphi_v (x - \bar{x})/2\sqrt{3}]$ a
dimensionless variable.

We may generate a semiclassical series by expanding the
$e^{-\delta S}$ term, and using the quadratic propagator in the
kink background to Wick-contract the various products of $\eta$
fields that will appear. Note that, when compared with perturbation
theory, this expansion will involve different vertices, as well
as a different propagator. The latter is already an infinite sum
of one-loop perturbative diagrams \cite{bsas,pre1}. We now proceed
to construct the propagator, a crucial step to obtain the series.


\subsection{The quadratic propagator}
\label{propagator}

The quadratic propagator of our semiclassical expansion must
satisfy  \cite{semi,bsas,pre1}
\begin{equation}
\label{PROP}
\left\{-\partial_{\tau}^2 -\bnab^2
+ [M^2 - \frac{3}{2}M^2{\rm sech}^{2}(\xi)]
\right\}G(\tau,\bx;\tau',\bx')=\delta(\tau-\tau')\delta^{d}(\bx-\bx')\;,
\end{equation}
as well as the boundary conditions
\begin{equation}
\label{BC}
G(0,\bx;\tau',\bx')=G(\beta,\bx;\tau',\bx')=0\;,
\end{equation}
since the fluctuations vanish at $\tau=0$, $\beta$. 

The spectrum of the differential operator in (\ref{PROP}) is
that of a Posch-Teller potential, which is known
exactly \cite{morsefesh}. Many applications involving kinklike
profiles \cite{interfaces,dwft,munster,bg,kinks} have made use of that explicit 
knowledge of the spectrum. Indeed, in some of them the quadratic propagator was
expressed as a sum over the various eigenmodes of the operator,
with the respective eigenvalues in the denominator. That, however,
leads to very lengthy calculations \cite{interfaces,munster},
especially if one goes beyond one-loop \cite{munster}. Therefore,
it is useful to derive a {\em closed exact} expression for the propagator
which is actually the result of that sum. We can do it
by realizing that, in transverse momentum space, the problem
can be reduced to finding the propagator for an ordinary
differential equation.

Fourier transforming the transverse coordinates, as well
as the Euclidean time, we obtain the ordinary differential equation
\begin{equation}
\label{ODE}
\left\{-\partial_{x}^2 + \omega_{n}^2+ \bk^2 + [M^2
- \frac{3}{2}M^2{\rm sech}^{2}(\xi)]\right\}\gtil(\kappa_n;x,x')
=\delta(x-x')\;,
\end{equation}
where $\omega_n=2\pi n/\beta$, and we have defined the
dimensionless vector $\bkappan\equiv \frac{2}{M}(\omega_n, \bk)$.
As we will see, $\gtil$ only depends on
$\kappa_n=|\bkappan|=\frac{2}{M} \sqrt{\omega_{n}^2+\bk^2}$.
The propagator $G$ is given by
\begin{equation}
\label{FG}
G(\tau,\bx;\tau',\bx')=\frac{1}{\beta}\sum_{n=-\infty}^{\infty}
\int \frac{d^{d-1}k}{(2\pi)^{d-1}}\,\gtil(\kappa_n;x,x')
e^{i\omega_{n}(\tau-\tau')} e^{i\bk\cdot(\br-\br')}\;.
\end{equation}
In terms of the dimensionless propagator
$\gatil\equiv (M/2)\gtil$, and of the variable $\xi$,
defined in (\ref{L2}), the equation reads
\begin{equation}
\label{ODE2}
\left\{-\partial_{\xi}^2 + 4 + \kappa_n^2 -6\,{\rm sech}^2(\xi)
\right\}\gatil(\kappa_n;\xi,\xi')=\delta(\xi-\xi')\;.
\end{equation}
We can solve it by finding two linearly independent solutions,
$\phi_1(\xi)$ and $\phi_2(\xi)$,
of the homogeneous equation \cite{semi,bsas,pre1}
\begin{equation}
\label{ODE3}
\left\{-\partial_{\xi}^2 + 4 + \kappa_n^2 -6\,{\rm sech}^2(\xi)
\right\}\phi_i(\xi)=0\;.
\end{equation}
The solutions are obtained in Appendix \ref{HYPERGEOMETRICS}:
\begin{equation}
\label{ETA1}
\phi_1(u)= {\left(\frac{u}{1-u}\right)}^{b_n/2}
\left[1-\frac{6}{b_n+1}u+\frac{12}{(b_n+1)(b_n+2)}u^2\right]\;,
\end{equation}
\begin{equation}
\label{ETA2}
\phi_2(u)={\left(\frac{1-u}{u}\right)}^{b_n/2}
\left[1+\frac{6}{b_n-1}u+\frac{12}{(b_n-1)(b_n-2)}u^2\right]\;,
\end{equation}
where $u=(1-\tanh\xi)/2$ and $b_n=\sqrt{4+\kappa_n^2}$.
Expressions (\ref{ETA1}) and (\ref{ETA2}) are well-defined for $\kappa_n\ne 0$, i.e.,
$b_n>2$. We will return to this point later on. The propagator
can now be constructed (see Appendix \ref{HYPERGEOMETRICS}), yielding
\begin{equation}
\label{GETAETA2}
\gatil(\kappa_n;u,u')=\frac{1}{2b_n}\{\phi_1(u)\,
\phi_2(u')\,\Theta(u'-u)+\phi_1(u')\,
\phi_2(u)\,\Theta(u-u')\}\;,
\end{equation}
where $\Theta(x)$ is the Heaviside stepfunction.
Note that the
factors outside the brackets in (\ref{ETA1}) and (\ref{ETA2})
will cancel in (\ref{GETAETA2}) whenever $u=u'$.

The propagator we have just constructed can only be defined
for $\kappa_n>0$, or $b_n>2$. As is well known \cite{dhn,morsefesh},
for $\kappa_0=0$ the fluctuation kernel in (\ref{ODE2}) admits a
zero eigenmode associated to translational symmetry, i.e., a zero
eigenvalue solution of (\ref{ODE3}) that vanishes at
$\xi\to \pm\infty$. Explicitly:
\begin{equation}
\label{ZERO}
\etil_0(u)=\frac{\sqrt{3}}{2}\,{\rm sech}^{2}[\xi(u)]=2\sqrt{3}\,u(1-u)\;.
\end{equation}
Consequently, we can only define a propagator in the subspace
orthogonal to that spanned by (\ref{ZERO}). Thus, we subtract
the contribution of that mode to obtain
\begin{equation}
\label{G'}
\gatil'(\kappa_n;u,u')=\gatil(\kappa_n;u,u')
-\frac{\etil_0(u)\etil_0(u')}{\kappa_n^2}\;.
\end{equation}
The divergent part of the limit as $\kappa_n\to 0$ of
$\gatil(\kappa_n;u,u')$ is exactly cancelled by the second
term on the right-hand side of (\ref{G'}), so that
$\gatil'(\kappa_n;u,u')$ is well defined and orthogonal to
the zero mode subspace in the limit $\kappa_n\to 0$:
\begin{equation}
\label{G'2}
\lim_{\kappa_n\to 0}\gatil'(\kappa_n;u,u')
=\frac{3}{4}\,u(1-u)u'(1-u')\;.
\end{equation}
%


\subsection{Determinants}
\label{determinants}

The leading term of the semiclassical expansion of the partition
function around a domain wall background is obtained by setting
$\delta S=0$ in expression (\ref{Z2}). The quadratic path-integral
requires collective coordinates for the zero-mode
subspace \cite{rajaraman,helgaser}. As we will see in Section \ref{collective}, 
the remaining integration,
in the subspace orthogonal to that eigenmode, leads to a fluctuation
determinant which will be computed from the quadratic semiclassical
propagator for arbitrary temperature $T$ and $d=1,2,3$.

We again Fourier transform the fluctuation fields in all
coordinates but the longitudinal one:
\begin{equation}
\label{FOUETA}
\eta(\tau,\bx)=\frac{1}{\beta}\sum_{n=-\infty}^{\infty}\int
\frac{d^{d-1}k}{(2\pi)^{d-1}}\, \widetilde{\eta}(\bkappan;x)\,
e^{i\omega_{n}\tau +i\bk\cdot\br}\;.
\end{equation}
The quadratic part of the action may be written as
\begin{equation}
\label{S2SUM}
S_2=\frac{1}{\beta}\sum_{n=-\infty}^{\infty}\int
\frac{d^{d-1}k}{(2\pi)^{d-1}} {\widetilde S}_{2}(\bkappan)\;,
\end{equation}
\begin{equation}
\label{S2KN}
{\widetilde S}_{2}(\bkappan)=\frac{1}{2}\int_{-\infty}^{\infty}
dx \,\etil(\bkappan;x)\,K(\kappa_n)\,\etil(\bkappan;x)\;.
\end{equation}
We denote by $K(\kappa_n)$ the differential operator in
(\ref{ODE}), and {\em formally} define its determinant as
\begin{equation}
\label{DELTAKN}
\Delta_{\kappa_n}^{-1/2}=[{\rm det}K(\kappa_n)]^{-1/2}=
{\cal N} \int \left[{\cal D}\etil \right]
e^{-{\widetilde S}_{2}(\bkappan)}\;,
\end{equation}
where the path integral sums over all fluctuations satisfying
$\etil(\bkappan;\pm \infty)=0$, and ${\cal N}$ is a normalization constant. Denoting by 
$|\etil_j \rangle$ the eigenmodes (Posch-Teller modes, which do not depend on $\bkappan$),
and by $\gamma_{j\kappa_n}^2$ the eigenvalues of
$K(\kappa_n)$, respectively, we have
\begin{equation}
\label{GKN}
K^{-1}(\kappa_n)=\sum_{j}\!\!\!\!\!\!\!\int
\frac{|\etil_j \rangle\langle\etil_j|}{\gamma_{j\kappa_n}^2}\;,
\end{equation}
\begin{equation}
\label{GAMMAKN}
\gamma_{j\kappa_n}^2=\frac{1}{4}M^2 \kappa_n^2+ \gamma_{j}^2\;,
\end{equation}
where $j$ denotes both discrete and continuum Posch-Teller
eigenstates, and $\gamma_{j}=\gamma_{j0}$ are their respective eigenvalues.
By convention, $j=0$ denotes the lowest of the eigenstates.
As we have already remarked, $\gamma_{00}^2=0$, which implies
$\Delta_{0}=0$. Therefore, when in the $\kappa_0=0$ subspace,
we must use a primed determinant $\Delta'_{0}$ which, by
definition, is the product of all but the lowest eigenvalue.

We may now relate ratios of primed determinants (i.e., with
the lowest eigenvalue excluded) to the semiclassical propagator
of the previous section:
\begin{equation}
\label{RATIO1}
\ln{\frac{\Delta'_{\kappa_n}}{\Delta'_0}}
= \int_{0}^{\kappa_n^2} ds\,
{\rm Tr}\, \gatil'(\sqrt{s})\;,
\end{equation}
where the trace is
$\int_{-\infty}^{\infty} d\xi\, \gatil'(\sqrt{s}; \xi, \xi)$.
We are ultimately interested in ratios of determinants in
the presence of the domain wall to free ones: for $\kappa_0=0$,
this is just given by the corresponding ratio for the
Posch-Teller potential in one-dimensional quantum mechanics,
$\ln(\Delta'_{0}/\Delta^F_{0})$, which excludes the zero mode;
for $\kappa_n\ne 0$, we may write
\begin{equation}
\label{RATIO2}
\ln{\frac{\Delta_{\kappa_n}}{\Delta^F_{\kappa_n}}}
=\ln\kappa_n^2 + \ln{\frac{\Delta'_{\kappa_n}}{\Delta'_0}}
-\ln{\frac{\Delta^F_{\kappa_n}}{\Delta^F_0}}
+\ln{\frac{\Delta'_0}{\Delta^F_0}}\;.
\end{equation}
The first term in (\ref{RATIO2}) simply restores the lowest
eigenvalue for $\kappa_n \ne 0$, whereas the second and the third
can be calculated from the knowledge of propagators: the one in the
domain wall background, and the free one. From the calculation outlined
in Appendix \ref{DETERMINANTS2}, we obtain
\begin{equation}
\label{3TERMO}
\ln{\frac{\Delta'_{\kappa_n}}{\Delta'_0}}
-\ln{\frac{\Delta^F_{\kappa_n}}
{\Delta^F_0}}=\ln\left(\frac{b_n-1}{b_n+1}\right)-2\ln(b_n+2)+\ln 48\;.
\end{equation}
We may follow \cite{rajaraman} to derive the result for
$\kappa_0=0$:
\begin{equation}
\label{2TERM}
\ln{\frac{\Delta'_{0}}{\Delta^F_0}}=-\ln 48\;.
\end{equation}
If we insert (\ref{3TERMO}) and (\ref{2TERM}) into (\ref{RATIO2}),
which is valid for $\kappa_n \ne 0$, we obtain
\begin{equation}
\label{RATIO3}
\ln{\frac{\Delta_{\kappa_n}}{\Delta^F_{\kappa_n}}}
=\ln\left(\frac{b_n-1}{b_n+1}\right)
+\ln\left(\frac{b_n-2}{b_n+2}\right)\;.
\end{equation}
The above result includes the $n=0$ subspace of eigenmodes,
but requires $\kappa_n \ne 0$, as the $\kappa_0=0$ contribution
has been explicitly summed to cancel the $\ln 48$ term of
(\ref{3TERMO}).

{\em Formally}, the logarithm of the complete (primed)
determinant is the sum over all possible $\bkappan $'s
($\kappa_n > 0$) of (\ref{RATIO3}):
\begin{equation}
\label{COMPDET}
\ln{\frac{\Delta'}{\Delta^F}}=V_{d-1}\int \frac{d^{d-1}k}
{(2\pi)^{d-1}}\,\sum_{n=-\infty}^{\infty}
\ln{\frac{\Delta_{\kappa_n}}{\Delta^F_{\kappa_n}}}\;.
\end{equation}

Expression (\ref{COMPDET}), as it stands, is meaningless:
the sum over $n$ diverges logarithmically, and, for $d>1$,
the integral over $\bk$ introduces extra divergences
as well. One must undertake a renormalization process in order
to arrive at well-defined results. In the next section, we
shall regularize, and then renormalize the calculation to
obtain a final answer for any $T$ and for $d=1,2,3$, where
the theory is renormalizable.


\section{ The Renormalization Procedure}
\label{renormalization}


\subsection{Regularization}
\label{regularization}

We shall concentrate our attention on (\ref{COMPDET}).
We begin by examining the behavior of the sum ${\cal S}$
over Matsubara frequencies defined as
\begin{equation}
\label{MATSUM1}
{\cal S}\equiv \sum_{n=-N}^{N}\, \ln{\frac{\Delta_{\kappa_n}}
{\Delta^F_{\kappa_n}}}\;,
\end{equation}
where $N$ is a cutoff. If we introduce a dimensionless
function $c_k\equiv (\beta/2\pi)\sqrt{M^2+k^2}$,
the sum becomes
\begin{equation}
\label{MATSUM2}
{\cal S}=\sum_{n=-N}^{N}\left(\ln\frac{\sqrt{n^2+c_k^2}
- \frac{1}{2}\,c_0}{\sqrt{n^2+c_k^2}+ \frac{1}{2}\,c_0}
+ \ln\frac{\sqrt{n^2+c_k^2}- c_0}
{\sqrt{n^2+c_k^2}+ c_0}\right).
\end{equation}
${\cal S}$ may be rewritten in a convenient way by resorting
to expression (\ref{PLANAFORMULA}) of Appendix \ref{PLANA},
obtained from Plana's formula \cite{ww}, which is commonly
used in the description of the Casimir effect \cite{mostep}.
For $\kappa_n \ne 0$, it is given as the difference
of two terms: ${\cal S}= {\cal S}_a - {\cal S}_b$, where
\begin{equation}
\label{MATSUM3}
{\cal S}_a=\sum_{j=1}^{2} \sum_{\sigma=0}^{1} 2\,
e^{i\sigma\pi}\left[\int_{0}^{N} dx \ln \left(\sqrt{x^2+c_k^2}
-\frac{1}{2}\, j e^{i\sigma\pi} c_0 \right)\right],
\end{equation}
\begin{equation}
\label{MATSUM4}
{\cal S}_b=\sum_{j=1}^{2}\int_{c_k}^{\infty}\frac{8\, dy}
{e^{2\pi y} - 1}\left[\pi - \arctan
\left(\frac{2\sqrt{y^2 - c_k^2}}{j c_0}\right)\right].
\end{equation}
The integral $I$ in the brackets of (\ref{MATSUM3})
can be done exactly [see (\ref{INTEGRAL}) in Appendix \ref{INTEG}].
For $N$ large, if we neglect terms of order $1/N$, replace $N$
with $\beta\Lambda/2\pi$, and restore dimensional quantities,
expression (\ref{MATSUM2}) can be split into a zero temperature
part ${\cal S}_0$,
\begin{equation}
\label{MATSUM0}
{\cal S}_0=\frac{2\beta k}{\pi} \arcsin{\left(\frac{c_0}
{c_k}\right)}+\frac{2\beta}{\pi}\sqrt{k^2+\frac{3}{4}M^2}
\arcsin{\left(\frac{c_0}{2c_k}\right)}-\frac{3}{\pi}
\beta M\left[\ln{\left(\frac{\beta\Lambda}{\pi c_k}\right)}
+ 1\right],
\end{equation}
and a temperature dependent part ${\cal S}_T$,
\begin{equation}
\label{MATSUMT}
{\cal S}_T=\int_{c_{k}}^{\infty}\frac{8\pi dy}{e^{2\pi y} -1}
\left[\arctan{\left(\frac{\sqrt{y^2-c_k^2}}{c_0}\right)}
+ \arctan{\left(\frac{2\sqrt{y^2-c_k^2}}{c_0}\right)}\right]
+ 4\ln{\left(1- e^{-2\pi c_k}\right)}\;.
\end{equation}
In both expressions, the first two terms represent the
contribution of the (Posch-Teller) bound states in the
longitudinal direction, while the last ones correspond to
the continuum. For $d=1$, $k=0$, so
that the first term of (\ref{MATSUM0}) vanishes.


\subsection{Renormalization}
\label{renorm}

The continuum contribution (last term) to the zero temperature
part ${\cal S}_0$ contains the expected logarithmic divergence.
Indeed, at zero temperature we should replace the sum
(\ref{MATSUM1}) by an integral:
\begin{equation}
\label{MATINT}
{\cal S}= \sum_{n=-N}^{N}\, \ln{\frac{\Delta_{\kappa_n}}
{\Delta^F_{\kappa_n}}}\, \stackrel{T\to 0}{\longrightarrow}\, 
{\cal I}= \beta\int_{-\infty}^{\infty} \frac{d\omega}{2\pi}
\ln{\frac{\Delta_{\kappa}}{\Delta^F_{\kappa}}}\;,
\end{equation}
where $\omega$ is a continuous variable that replaces
$\omega_n$, and $\bkappa\equiv \frac{2}{M}(\omega,\bk)$.
Furthermore, we may write
\begin{equation}
\label{RATINT}
\ln{\frac{\Delta_{\kappa}}{\Delta^F_{\kappa}}}
={\rm Tr}\ln\left[\openone - \gatil_F(\kappa)U\right]\;,
\end{equation}
where $\openone$ is the identity operator, $\gatil_F(\kappa;\xi,\xi')$
is given by (\ref{FREEPROP}), with $b_n$ replaced by $b\equiv\sqrt{4+\kappa^2}$,
$U(\xi)\equiv -6\,{\rm sech}^{2}(\xi)$ is the interaction term in
(\ref{ODE2}), and the trace is defined in (\ref{RATIO1}).
If we expand the logarithm in (\ref{RATINT}), we shall have
a series of one-loop terms:
\begin{equation}
\label{ONELOOP}
\ln{\frac{\Delta_{\kappa}}{\Delta^F_{\kappa}}}
=-\sum_{n=1}^{\infty} \frac{1}{n}\,{\rm Tr}\left[
\gatil_F(\kappa)U\right]^{n}.
\end{equation}
Inserting (\ref{ONELOOP}) into (\ref{MATINT}), the $n=1$
term ${\cal I}_1$ can be readily computed (change the
integration variable from $\xi$ to $u$)
and identified with the divergent term of (\ref{MATSUM0}):
\begin{equation}
\label{N=1}
{\cal I}_1=-\beta \int
\frac{d\omega}{2\pi} \int_{-\infty}^{\infty}\!\!d\xi\,
\gatil_F(\kappa;\xi,\xi)\,U(\xi)=-\,\frac{3}{\pi}
\beta M\ln{\left(\frac{\beta\Lambda}{\pi c_k}\right)}\;.
\end{equation}
In order to renormalize the theory, we add a counterterm to the
Lagrangian which will absorb that divergent term, and impose
renormalization conditions to fix the finite remainder. For simplicity,
we will adopt renormalization conditions at zero momentum $\bkappa$. 
In the present case, it suffices to add a mass-type counterterm:
\begin{equation}
\label{CT1}
{\cal L}[\varphi]\to {\cal L}[\varphi]- \frac{1}{2}C_1(\varphi^2-\varphi_v^2)\;,
\end{equation}
and to require that the one-point one-particle irreducible (1PI) 
Green function at zero momenta $\Gamma_R^{(1)}$ be given by:
\begin{equation}
\label{RN1}
{\widetilde\Gamma}_R^{(1)}({\bf{0}})=0\;.
\end{equation}
In order to satisfy (\ref{RN1}), we must subtract the tadpole
graph in the kink background, which is also given by (\ref{N=1})
because the free propagator at coincident points can be factored out
of the $\xi$ integration. We then obtain the renormalized expression
\begin{equation}
\label{MS1}
{\cal S}_{{\rm R}}=\frac{2 \beta k}{\pi}
\arcsin{\left(\frac{c_0}{c_k}\right)}+\frac{2\beta}{\pi}
\sqrt{k^2+\frac{3}{4}M^2}\arcsin{\left(\frac{c_0}{2c_k}\right)}
-\frac{3}{\pi}\beta M + {\cal S}_T\;.
\end{equation}
For $d=1$, $k=0$, $c_k=c_0$, and there are no additional
integrals. In that case, we may verify that the temperature independent
part of (\ref{MS1}) coincides with the result quoted in equation (3.29) 
of the first reference in \cite{rajaraman}. Indeed, from our calculation
we may easily obtain the difference between the energy of the kink and 
that of the vacuum, i.e., the kink mass:
\begin{equation}
\label{kinkmass1}
M_{kink}=\frac{1}{\beta}\left(\frac{S[\fith]}{\lambda}+\frac{1}{2}
{\cal S}_{{\rm R}}\right)\;,
\end{equation}
which at zero temperature yields:
\begin{equation}
\label{kinkmass2}
M_{kink}=\frac{2M^3}{\lambda}+ \frac{M}{\sqrt{2}}\left(
\frac{1}{2\sqrt{6}}-\frac{3}{\pi\sqrt{2}}\right)\;.
\end{equation}
The replacements $M=\sqrt{2}m_{raj}$, and $\lambda=6\lambda_{raj}$, needed to account
for different definitions of mass and coupling, show that our results coincide with 
Rajaraman's \cite{rajaraman} for $d=1$. We note that (\ref{3TERMO}) and (\ref{2TERM}) had already made use of that coincidence for the case $d=0$. 

For $d=2$, and $d=3$, we still have to
integrate over $\bk$. This last integral will
generate additional divergences in $d=3$ dimensions, which
are contained in the $n=2$ term of (\ref{ONELOOP}):
\begin{equation}
\label{N=2}
{\cal I}_2=V_{d-1}\int \frac{d^{d-1}k}{(2\pi)^{d-1}}
\beta\int_{-\infty}^{\infty} \frac{d\omega}{2\pi}
{\cal J}_2\;,
\end{equation}
\begin{equation}
\label{J2ff}
{\cal J}_2\equiv -\,\frac{1}{2}\int_{-\infty}^{\infty}\!\!d\xi_1\,
\int_{-\infty}^{\infty}\!\!d\xi_2\,\gatil_F(\kappa;\xi_1,\xi_2)\,
U(\xi_2)\,\gatil_F(\kappa;\xi_2,\xi_1)\,U(\xi_1)\;.
\end{equation}
The calculation of Appendix \ref{SECONDORDER} leads to
\begin{equation}
\label{J2f}
{\cal J}_2=- 36\left\{\frac{1}{2b^2(b+1)^2}
+\frac{1}{6(b+1)^3}+ O[1/(b+1)^5]\right\}.
\end{equation}
Since $b= \sqrt{4+\kappa^2}$,
the second term in (\ref{J2f}) dominates in the ultraviolet.

We now analyze expression (\ref{N=2}) for each dimension:
i) in $d=1$, the integral over $\omega$ goes like
$\int\omega^{-3}\, d\omega\sim \Lambda^{-2}$, which vanishes
as $\Lambda \to \infty$; ii) in $d=2$, the integrals over
$\omega$ and $k$ lead to ${\cal I}_2\sim \Lambda^{-1}$,
which also vanishes in that limit; iii) in $d=3$, however,
the integrals over $\omega$ and $k$ lead to
${\cal I}_2\sim \ln \Lambda$, which diverges in that limit.
Thus, in $d=3$ an additional subtraction is required. The leading 
ultraviolet behavior of (\ref{N=2}) comes from the second term in (\ref{J2f}):
\begin{equation}
\label{I2}
{\cal I}_2^{({\rm as})}=V_{2}\int \frac{d^{2}k}{(2\pi)^{2}}\,\beta
\int_{-\infty}^{\infty} \frac{d\omega}{2\pi}
\frac{-6(M/2)^3}{\left[\sqrt{\omega^2+k^2+M^2}+
(M/2)\right]^3} \;.
\end{equation}
We integrate the momenta $\bk$ over a sphere of radius
$\Lambda$ to obtain \cite{grad}
\begin{equation}
\label{I2UV}
{\cal I}_2^{({\rm as})}=-\frac{3}{16\pi^2}\,V_2\beta M^3
\left[\ln(\Lambda^2/M^2)+ O(1)\right].
\end{equation}
This is to be compared with the result of integrating
the finite terms of (\ref{MATSUM0}) over  momenta
$\bk$ (the divergent term was already cancelled by
the firts counterterm). Integration by parts leads to elementary
integrals; expanding the result for large $\Lambda$,
and neglecting terms of order $1/\Lambda^2$, we obtain
\begin{equation}
\label{D2KS0}
V_2\int \frac{d^{2}k}{(2\pi)^{2}}\,{\cal S}_0
=-\frac{3}{16\pi^2}\,V_2\beta M^3\,\ln(\Lambda^2/M^2)
-\frac{\sqrt{3}}{48 \pi}\,V_2\beta M^3 + O(1/\Lambda^2)\;.
\end{equation}
As expected, the leading ultraviolet behavior is the same, which shows that
the $n=2$ term of (\ref{ONELOOP}) is indeed responsible for the new divergence.
The renormalization procedure requires that we include an additional counterterm in
the Lagrangian, and use an additional renormalization condition to fix finite
parts in $d=3$ (see Appendix \ref{EFFPOT}). The Lagrangian becomes:
\begin{equation}
\label{CT2}
{\cal L}[\varphi]\to {\cal L}[\varphi]- \frac{1}{2}C_1(\varphi^2-\varphi_v^2)-
\frac{1}{4}C_2(\varphi^2-\varphi_v^2)^2\;.
\end{equation}
$C_1$ and $C_2$ are counterterms that we fix by imposing (\ref{RN1}), as well as:
\begin{equation}
\label{RN2}
{\widetilde\Gamma}_R^{(2)}({\bf{0}},{\bf{0}})=M^2\;.
\end{equation}
The contribution of the additional counterterm to the effective action in the
kink sector is given by:
\begin{equation}
\label{CT3}
\left[-\frac{1}{2}\int \frac{d^3 k}{(2\pi)^3}\int^{\infty}_{-\infty} \frac{dw}{2\pi}\frac{1}{(w^2+k^2+M^2)^2}\right] 
\left[\frac{V_2}{4}\int^\beta_0 d\tau \int_{-\infty}^{\infty} dx\,(\varphi^2-\varphi_v^2)^2
\right]\;,
\end{equation}
where the first bracket is identified as the one-loop graph with two external legs at zero momenta (integrated over $w$ in the interval $[-\infty,\infty]$, and over $\bk$ in the two-sphere of radius $\Lambda$, as before), while the second is easily related to the classical kink action. A simple calculation yields:
\begin{equation}
\label{CT4}
-\frac{3}{16\pi^2}\,V_2\beta M^3\,\ln(\Lambda^2/M^2)
+\frac{3}{16 \pi^2}\,V_2\beta M^3 \;.
\end{equation}
If we subtract the result of (\ref{CT4}) from (\ref{D2KS0}), and take the limit $\Lambda
\to \infty$, we finally arrive at  
\begin{equation}
\label{DETREN}
\ln{\left(\frac{\Delta'}{\Delta^F}\right)_{\rm R}}
=-\frac{3}{16 \pi^2}\left[1+ \frac{\pi}{3\sqrt{3}}\right]\,V_2\beta M^3+
V_2 \int \frac{d^{2}k}{(2\pi)^{2}}\,{\cal S}_T\;.
\end{equation}

We should point out that expressions for the finite temperature determinant were
previously obtained from the explicit knowlegde of the Posch-Teller eigenvalues,
either by summing over Matsubara frequencies, and then over those eigenvalues 
\cite{dwft}, or by integrating over transverse momenta in arbitrary dimension first,
making use of dimensional regularization, and then summing over Matsubara frequencies 
\cite{bg}. 
\section{Collective Coordinates}
\label{collective}

The domain wall solution breaks translational symmetry along the longitudinal 
$x$-axis, as it depends on a parameter $\barx$ that characterizes its position. 
In order to restore that symmetry of the theory, we resort to the method of 
collective coordinates \cite{rajaraman} to sum over all possible values of $\barx$.
In $d=1$, there is nothing else to restore. In $d>1$, where the wall also breaks rotational
invariance by singling out a normal direction, consistently with our
previous assumptions, the longitudinal direction is externally fixed by the forces
that implement the wall boundary conditions at spatial infinity. Therefore, here
as well, longitudinal translations are the only degrees of freedom to consider.

The existence of longitudinal translation zero eigenmodes of the fluctuation operator 
is a clear signal that the symmetry has to be restored. Physically, it means that it 
costs no energy to go from one solution to a longitudinally translated one. Thus, in
the usual manner, in a region of longitudinal length $L$, we introduce in the partition 
function the identity
\begin{equation}
\label{identity}
\int \frac{d\barx}{L}\, \delta\left(\frac{1}{L \lambda^{1/2}}\int_0^{\beta} d\tau \int d^d x\,\eta_0(x-\barx)
\left[\varphi(\tau,x,\br)-\widehat\varphi(x-\barx)\right]\right)J[\varphi]=1\,,
\end{equation}
where the Jacobian is given by
\begin{equation}
\label{jacobian}
J[\varphi]=\lambda^{-1/2}\int_0^{\beta} d\tau \int d^d x\, \varphi(\tau,x,\br) 
\frac{\partial}{\partial \barx}
\eta_0(x-\barx)=\lambda^{-1/2}\int_0^{\beta} d\tau \int d^d x\, \eta_0(x-\barx) 
\frac{\partial\varphi}{\partial x}\,.
\end{equation}
Equation (\ref{identity}) imposes orthogonality to the normalized zero
mode $\eta_0$, and integrates 
over longitudinal positions. The second equality in (\ref{jacobian})
comes from an integration by parts. We note that the zero mode is 
already orthogonal to the domain
wall profile since $\eta_0\propto \partial{\widehat\varphi}/\partial x$, 
which implies
\begin{equation}
\label{R=0}
\int_{-\infty}^{\infty} dx\, \eta_0(x-\barx)\widehat\varphi(x-\barx)\propto 
\left[{\widehat\varphi}^2
\right]_{-\infty}^{\infty}=0\,.
\end{equation}
Therefore, we may omit the $\widehat\varphi$ term in (\ref{identity}). 
We insert the identity in
the expression for the partition function (\ref{Z1}), and change the 
order of integration to obtain
\begin{equation}
\label{Zmode1}
Z(\beta)=\int \frac{d\barx}{L}\,\oint
[{\cal D}\varphi]\, e^{-S[\varphi]/\lambda}\,\delta\left(\frac{1}{L \lambda^{1/2}}\int_0^{\beta} 
d\tau \int d^d x\, 
\eta_0(x-\barx) \varphi(\tau,x,\br)\right)J[\varphi]\,.
\end{equation}
We perform the semiclassical expansion around the domain wall 
according to (\ref{expansion}). Since
the action can be written as
\begin{equation}
\label{action}
S[\fith]=\int_0^{\beta} d\tau \int d^d x\, 
\left(\frac{\partial\fith}{\partial x}\right)^2,
\end{equation}
the normalized zero eigenmode is simply 
$\eta_0={S[\fith]}^{-1/2}(\partial\fith/\partial x)$. Using
that, and its orthogonality to $\fith$, the expansion yields
\begin{equation}
\label{Zmode2}
Z(\beta)=\int \frac{d\barx}{L}\, e^{-S[\fith]/\lambda}\,\oint
[{\cal D}\eta]\, e^{-S_2}\,\delta\left(\frac{1}{L}\int_0^{\beta} 
d\tau \int d^d x\, \eta_0
\eta\right)\,J[\fith,\eta]\,\sum_{n=0}^\infty \frac{1}{n!}\,(-\delta S)^n\,,
\end{equation}
with the functional integral being performed over fluctuations that 
vanish at $\tau=0,\beta$, 
and the expanded Jacobian given by
\begin{equation}
\label{jacobexp}
J[\fith,\eta]=\left(\frac{S[\fith]}{\lambda}\right)^{1/2} - \int_0^{\beta} d\tau \int d^d x\, 
\frac{\partial\eta_0} {\partial x} \eta = \left(\frac{S[\fith]}{\lambda}\right)^{1/2}-
{S[\fith]}^{-1/2}\int_0^{\beta} d\tau \int d^d x\, V'[\fith]\, \eta\,.
\end{equation}

The leading term in (\ref{Zmode2}) corresponds to the quadratic approximation $Z_2$
to the partition function in the domain wall background, whose ratio to the free one
is given by 
\begin{equation}
\label{Zquad}
\frac{Z_2(\beta)}{Z_F}=\int \frac{d\barx}{L}\, e^{-S[\fith]/\lambda}\, 
\left[\frac{\Delta'}{\Delta_F}
\right]_{\rm R}^{-1/2}\, \left(\frac{S[\fith]}{2\pi\lambda}\right)^{1/2}\,.
\end{equation}
It involves the primed determinant calculated previously, since 
the delta function restricts us 
to the subspace orthogonal to the zero eigenmode, has a factor $(S[\fith]/\lambda)^{1/2}$ 
coming from 
the Jacobian, and a factor of ${(2\pi)}^{-1/2}$ from the functional measure 
which is not cancelled 
by the corresponding value for the free determinant in the zero mode subspace.

\section{Physical Applications and Higher Orders}
\label{physappl}

The result obtained in the preceding section has an immediate 
physical application in the
calculation of the interface tension in $d=2$ and $d=3$. 
If our system is confined to a longitudinal
length $L$, and a transverse cross-section $V_{d-1}$, the free 
energy difference per unit cross-section
between the situation with and without the domain wall defines the interface 
(surface) tension \cite{interfaces}
\begin{equation}
\label{tension1}
\sigma\equiv - \lim_{V_{d-1}\to\infty}\frac{1}{\beta V_{d-1}}
\lim_{L\to\infty}\ln\left(\frac{Z}{Z_F}\right)\,,
\end{equation}
where $Z$ denotes the partition function in the presence of the wall, 
and $Z_F$ the free partition function, 
without the domain wall. The lowest order result (\ref{Zquad}), 
with the integral over $\bar{x}$ taken from $-L/2$
to $L/2$ gives
\begin{equation}
\label{tension2}
\sigma=\lim_{V_{d-1}\to\infty}\frac{1}{\beta V_{d-1}}
\left\{\frac{S[\fith]}{\lambda}+\frac{1}{2}
\ln\left(\frac{\Delta'}{\Delta_F}\right)_{\rm R}\right\}\,,
\end{equation}
where we have left out terms vanishing as $\ln V_{d-1}/V_{d-1}$ 
or faster. Another way of deriving the
interface tension is through its relationship with the energy 
splitting between ground and first excited 
state of the field theory, when its vacuum degeneracy is broken 
by the finite volume \cite{domb}. The
tunneling between the previously degenerate states can be 
computed using domain walls \cite{fisher,privman}.
This calculation was performed \cite{munster0} using a dilute 
gas of kinks and antikinks \cite{coleman}, and
yields
\begin{equation}
\label{tension3}
\Delta E= 2 e^{-S[\fith]/\lambda}\, \left[\frac{\Delta'}{\Delta_F}
\right]_{\rm R}^{-1/2}\, \left(\frac{S[\fith]}{2\pi\lambda}\right)^{1/2},
\end{equation}
an expression that comes from the exponentiation of (\ref{Zquad}), which results from
summing over kinks and antikinks. Our results incorporate finite temperature
effects into the expressions obtained in Ref.\cite{munster0}. 
Again, we stress that our starting
point is a field theory that has no phenomenological temperature 
dependence in its Lagrangian,
unlike treatments that start from a phenomenological Ginzburg-Landau 
coarse grained free energy
\cite{interfaces,munster}. Clearly, for $d=1$, a case that can be 
applied to the study of soliton
backgrounds in one-dimensional polymers, we would be comparing 
the free energy difference with and
without the soliton background ($V_0=1$), and finding how it is 
affected by thermal phonons.

The result obtained for the surface tension also provides us 
with an indirect determination
of the temperature of the second order transition that will 
restore the $\varphi\to-\varphi$
symmentry. Strictly speaking, it determines the so-called 
``percolation temperature'', 
the temperature at which the surface tension vanishes, which 
coincides with the
critical temperature in $d=3$, although it is smaller in $d=2$ 
\cite{lebowitz}. In fact, the 
statistical mechanics literature just compares free energies 
with and without domain wall
boundary conditions, and establishes that the signal for the 
symmetry restoring transition
is given when these two are equal. That means that the system 
will prefer to form walls
separating regions at different vacua, whose condensation leads 
to a vanishing order paramenter.
Our calculation is just the semiclassical version of that. 
That estimate has already been made
long ago \cite{dwft}, using a different method, as already explained.

Another physical information we can extract from our 
calculations is how the soliton ($d=1$), or 
interface ($d=2,3$) profiles are modified by thermal and 
quantum fluctuations. In lowest order, we may
use the one-loop expression for the effective action,
\begin{equation}
\label{effaction}
{\cal A}[\langle\varphi\rangle]=\frac{S[\langle\varphi\rangle]}{\lambda}
+\frac{1}{2}\ln( \Delta[\langle\varphi\rangle])\,,
\end{equation}
where $\langle\varphi(\bx)\rangle$ is the expectation value
 of the field, and ${\cal A}$ is the Legendre
transform of the free energy. We should look for extrema of 
(\ref{effaction}), so we functionally differentiate
with respect to $\langle\varphi(\bx)\rangle$. The resulting 
equation will just be the classical equation of 
motion plus a one-loop correction which involves the propagator 
in the $\langle\varphi(\bx)\rangle$ background.
Neglecting the one-loop part, we know that the domain wall 
profile is a solution. Thus, the ansatz
\begin{equation}
\label{ansatz}
\langle\varphi(\bx)\rangle=\fith(x)
+ \lambda^{1/2} \delta\langle\varphi(\bx)\rangle\,,
\end{equation}
leads to an equation for the correction term that 
involves the propagator in the domain wall background,  
for which we have an explicit expression. As a result, 
we will see how the thermal and quantum fluctuations
change the profile, and should be able to compare that 
prediction with scattering data from radiation
(of appropriate wavelength) scattered off the domain wall.

We close this section with comments about higher order 
corrections. In our construction, these are
generated by expanding the $e^{-\delta S}$ term in (\ref{Z2}), 
and using the quadratic propagator in the
domain wall background to Wick-contract the various 
products of $\eta$ fields that will appear. There is
also an additional vertex that comes from the second 
term of the Jacobian in (\ref{jacobexp}), which is 
linear in $\eta$, and proportional to $\lambda^{1/2}$. 
The first correction to the lowest order result
for the partition functions involves graphs which were 
computed in the second article of Ref.\ \cite{munster},
and represent a two-loop correction. Those same graphs 
can now be calculated with our propagator either
at $T=0$ or at finite temparature ($\tau$ integrals going 
from $0$ to $\beta$). In order to check and
extend those results, one has to use the same renormalization 
scheme, conveniently chosen to facilitate
comparisons with Monte Carlo data \cite{munster}. 

Numerical calculations will be required in applying 
even our lowest order results, as well as to allow
for comparisons with experimental or Monte Carlo data. 
The two-loop calculation just mentioned will also
require a numerical effort. We plan to undertake such 
efforts in the near future.   


\section{Conclusions}
\label{conclusions}

We have shown how one can systematically compute quantum fluctuations around
a domain wall background which is temperature independent, but whose fluctuations
interact with a heat bath at temperature $T$. The fluctuation determinant, the 
leading term in our semiclassical expansion, already makes the effective action 
$T$-dependent, and one can obtain an extremum for it that corresponds to a dressed 
$T$-dependent version of the original domain wall. 
As higher orders are included, that dependence will obviously change. The
determinant also leads to the calculation of the surface tension, which again 
will depend on $T$ through the fluctuations. In both cases, one can
follow the computations reviewed in Ref.\ \cite{interfaces}, omitting their
$T_{\rm ph}$-dependence in the paramenters of the Lagrangian, and replacing the 
determinants with our $T$-dependent ones. Also, the temperature
$T$ at which the surface tension vanishes can be used to estimate the phase
transition to the unbroken phase \cite{dwft}.

As we have already mentioned, results for $d=1$ can be applied to describe how
thermal phonons affect soliton backgrounds in polymer physics (examples involving
magnetic materials are also potential applications), 
while those for $d=2,3$ are applicable to Bose systems which are forced to condense 
in different phases in two different regions separated by a domain wall. The simple 
analytic form of the propagator in the domain wall background is quite useful in
calculations, and should lead to the determination of physical quantities 
measured in scattering experiments. 

In principle, the methods we use could be applied to other
theories, as long as the background solutions are sufficiently
symmetric so as to depend on only one spatial variable (a radial
one, for instance). The basic ingredient required is the knowledge,
exact or approximate, of two linearly independent solutions
of the fluctuation ODE at zero eigenvalue, from which we
construct the semiclassical propagator. Theories with solitonic (kinks,
vortices, monopoles, etc.) or instantonic backgrounds often
have such a property.

We hope, in the very near future, to present detailed numerical calculations
of the various quantities mentioned in Section \ref{physappl}. From them, we
expect to establish how realistic our description is, by confronting it with
available experimental data.


\acknowledgements

The author acknowledges support from CNPq, CAPES, and FUJB/UFRJ. This work was 
initiated during an extended visit to the Physics Department at UCLA, and 
finished during a short visit to the ICTP, in Trieste. It is a pleasure to
thank both institutions for their support and hospitality. Thanks are also
due to R. M. Cavalcanti for a careful reading of the manuscript.


\appendix{}

\section{}
\label{HYPERGEOMETRICS}

In this Appendix, we compute the solutions to (\ref{ODE3}), and use them to
construct the propagator (\ref{GETAETA2}). The standard substitution \cite{morsefesh}
$\phi(\xi)=e^{-a\xi}\cosh^{-b_n}(\xi){\rm F}(\xi)$,
with $a=0$ and $b_n=\sqrt{4+\kappa_n^2}$, and the change of
variable $u=(1-\tanh \xi)/2$ lead to a hypergeometric
equation for ${\rm F}$:
\begin{equation}
\label{HYPER}
u(1-u)\frac{d^2 {\rm F}}{du^2}+[(b_n+1)-2(b_n+1)u]
\frac{d{\rm F}}{du}+[6-b_n(b_n+1)]{\rm F}=0\;.
\end{equation}
The general solution is the linear combination
\begin{equation}
\label{GSOL}
{\rm F}(u)=c_1\,{}_2F_1(b_n-2,b_n+3; 1+b_n; u)
+c_2\, u^{-b_n}\,{}_2F_1(3,-2; 1-b_n; u)\;,
\end{equation}
where ${}_2F_1(A,B; C; u)$ is the hypergeometric function.
Using the identity \cite{grad}
\begin{equation}
\label{ID}
{}_2F_1(A,B; C; u)=(1-u)^{C-A-B}{}_2F_1(C-A,C-B; C; u)\;,
\end{equation}
we obtain
\begin{equation}
{\rm F}(u)=c_1\, (1-u)^{-b_n}\,{}_2F_1(3,-2; 1+b_n; u)
+c_2\, u^{-b_n}\,{}_2F_1(3,-2; 1-b_n; u)\;.
\end{equation}
Both series terminate, and we are led to the solutions in (\ref{ETA1}) and
(\ref{ETA2}). The propagator
can now be constructed from the functions
\begin{equation}
\label{OMEGA}
\Omega(\xi,\xi')\equiv \phi_1(\xi)\phi_2(\xi')
-\phi_1(\xi')\phi_2(\xi)\;,
\end{equation}
\begin{equation}
\label{W12}
W(\xi)\equiv\phi_1(\xi)\phi'_2(\xi)-\phi'_1(\xi)\phi_2(\xi)\;.
\end{equation}
%
It is given by \cite{semi,bsas,pre1}
\begin{equation}
\label{OMW}
\gatil(\kappa_n;\xi,\xi')=\frac{\Omega(\infty,\xi_{>})\,
\Omega(\xi_{<},-\infty)}{W(\xi)\,\Omega(-\infty,\infty)}\;,
\end{equation}
where $\xi_{<(>)}\equiv {\rm min(max)}(\xi,\xi')$. In
terms of the variable $u$, noting that $\xi>\xi'\Leftrightarrow u<u'$,
$\xi\to\infty \Rightarrow u\to 0$, and
$\xi\to -\infty \Rightarrow u\to 1$, we obtain
\begin{equation}
\label{GETAETA}
\gatil(\kappa_n;u,u')=\frac{\phi_1(u_{<})\,
\phi_2(u_{>})}{W(u)}\;.
\end{equation}
The propagator does vanish when $u,u'= 0,1$. Also, $W(u)$
is a constant, easy to compute at $u=0$; we find $W(u)=2b_n$.
We can then write expression (\ref{GETAETA2}).


\section{}
\label{DETERMINANTS2}

In this Appendix, we evaluate (\ref{3TERMO}) from the propagators. The free 
one, which can also be obtained from the methods described previously, is 
given by
\begin{equation}
\label{FREEPROP}
\gatil_F (\kappa_n;\xi,\xi')=\frac{1}{2b_n}\,e^{-b_n|\xi -\xi'|}\;.
\end{equation}
Changing variables to $b=\sqrt{4+s}$ and $u=(1-\tanh\xi)/2$,
the second term in (\ref{RATIO2}) becomes
\begin{equation}
\label{3TERM}
\ln{\frac{\Delta'_{\kappa_n}}{\Delta'_0}}=
\int_{2}^{b_n} db\, b \int_{0}^{1}\frac{du}{u(1-u)}
\left[\gatil'(\sqrt{b^2-4};u,u)-\gatil_F(\sqrt{b^2-4};u,u)\right].
\end{equation}
The expression inside the brackets is easily computed from
the propagators, yielding
\begin{equation}
\label{BRACKET}
\frac{6u(1-u)}{b(b^2-1)}-\frac{12(b^2+2b+3)u^2{(1-u)}^2}
{b(b+2)(b^2-1)}\;.
\end{equation}
Doing the integrals in (\ref{3TERM}), we arrive at (\ref{3TERMO}). 


\section{}
\label{PLANA}

In this Appendix, we shall make use of Plana's
formula \cite{ww,bateman} to compute the sum
(\ref{MATSUM2}), which can be decomposed into sums of the form
\begin{equation}
\label{ASUM}
{\cal S}(r,e^{i\sigma\pi} s)\equiv \sum_{n=-N}^{N}
\ln \left(\sqrt{n^2+r^2}+e^{i\sigma\pi} s\right)\;,
\end{equation}
with $r>s>0$, and $\sigma=0,1$.

Plana's formula can be derived from Cauchy's theorem applied
to two contours in the complex plane \cite{ww}: one that runs
counterclockwise in the upper half-plane, from $(x,y)=(n_1,n_1+i \infty)$  
to the real axis along a perpendicular, then along the real axis,
and finally perpendicularly away from it to 
$(n_2, n_2+ i \infty)$, avoiding the integers 
on the real segment $(n_1, n_2)$ (semicircling them), as well as the
corners $(n_1,0)$ and $(n_2,0)$ (with $\pi/2$ arcs); the
other is its mirror reflection on the real axis. For a function
$\phi(z)$ analytic and bounded for
$n_1\leq {\rm Re}(z)\leq n_2$, we have
\begin{equation}
\label{PLANA1}
\frac{\phi(n_1)}{2}+ \sum_{n=n_1+1}^{n_2-1} \phi(n)+
\frac{\phi(n_1)}{2}=\int_{n_1}^{n_2} dx\, \phi(x)+
\frac{1}{i}\int_0^{\infty}dy\, \frac{\Phi(n_2,y)-
\Phi(n_1,y)}{e^{2\pi y} - 1}\;,
\end{equation}
where $\Phi(n,y)\equiv \phi(n+iy)-\phi(n-iy)$. Defining
$\phi(z)= \ln \left(\sqrt{z^2+r^2}+e^{i\sigma\pi} s\right)$,
we may rewrite (\ref{ASUM}) as
\begin{equation}
\label{ASUM2}
{\cal S}(r,e^{i\sigma\pi} s)=2\sum_{n=0}^{N}\ln
\left(\sqrt{n^2+r^2}+e^{i\sigma\pi} s\right)-
\ln \left(r+ e^{i\sigma\pi}s\right),
\end{equation}
and use (\ref{PLANA1}) with $n_1=0^+$ (in order to avoid the
square-root cut at ${\rm Re}(z)=0$) and $n_2=N$. The contribution
to the second integral of (\ref{PLANA1}) involving $\Phi(N,y)$
can be expanded for large $N$, and shown to behave as $1/N$.
The one involving $\Phi(0^+,y)$ has to be split into two pieces:
$|y|> r$ and $|y|< r$. The latter piece yields zero, whereas
the former uses
\begin{equation}
\label{CUT}
\phi(0^+\pm iy)= \frac{1}{2}\ln [y^2-(r^2-s^2)] \pm
i\left[ \sigma\pi + e^{i\sigma\pi}\arctan\left(
\frac{\sqrt{y^2-r^2}}{s}\right)\right]\;.
\end{equation}
Neglecting terms that vanish as $N\to\infty$, we finally arrive at
\begin{eqnarray}
\label{PLANAFORMULA}
{\cal S}(r,e^{i\sigma\pi} s)&=& 2\int_{0}^{N} dx
\ln \left({\sqrt{x^2+r^2}+ e^{i\sigma\pi} s }\right)+
\ln \left({\sqrt{N^2+r^2}+ e^{i\sigma\pi} s }\right)
\nonumber \\
& &-4\int_r^{\infty} \frac{dy}{e^{2\pi y} - 1}\left[
\sigma\pi + e^{i\sigma\pi} \arctan \left(
\frac{\sqrt{y^2 - r^2}}{s}\right)\right]\;,
\end{eqnarray}
where $r>s>0$, and $\sigma=0,1$. It is easy to verify that
(\ref{PLANAFORMULA}) correctly reproduces the results for
${\cal S}(r,0)$ and for the sum
$\sum_{\sigma=0}^1{\cal S}(r,e^{i\sigma\pi} s)$,
which can be obtained straightforwardly.


\section{}
\label{INTEG}

In this Appendix, we calculate the integral $I$ that
appears inside the brackets of (\ref{MATSUM3}):
\begin{equation}
\label{INTEGR}
I(r,e^{i\sigma\pi} s)=\int_{0}^{N} dx
\ln \left(\sqrt{x^2+r^2}+e^{i\sigma\pi} s \right)\;.
\end{equation}
Changing variable to $q(x)=\sqrt{x^2+r^2}+ e^{i\sigma\pi} s$,
we integrate by parts to obtain
\begin{equation}
\label{INTEGRA}
I=\left[\sqrt{\left(q- e^{i\sigma\pi} s\right)^2 - r^2}
\,\,\ln q \right]_{q(0)}^{q(N)}- \int_{q(0)}^{q(N)}\frac{dq}{q}
\sqrt{\left(q- e^{i\sigma\pi} s\right)^2 - r^2}\;.
\end{equation}
The last integral can be found in \cite{grad}. Finally:
\begin{eqnarray}
\label{INTEGRAL}
I&=&N\ln\left(\sqrt{N^2+r^2}+ e^{i\sigma\pi} s \right)
- N  + e^{i\sigma\pi} s \left[ \ln \left(\frac{\sqrt{N^2+r^2}+N}
{r}\right)\right]
\nonumber \\
& &-\sqrt{r^2 - s^2}\left\{\arcsin{\left[\frac{r^2+ e^{i\sigma\pi}
s \sqrt{N^2+r^2}}{r\left(\sqrt{N^2+r^2}+ e^{i\sigma\pi} s \right)}
\right]}- \frac{\pi}{2}\right\}.
\end{eqnarray}
%


\section{}
\label{SECONDORDER}

We now compute the integral ${\cal J}_2$, defined in
(\ref{J2ff}). Again resorting to the change
of variables $\xi \to u $,
\begin{equation}
\label{J2a}
{\cal J}_2=\frac{- 36}{b^2}\int_0^1 du_1\left(\frac{u_1}
{1- u_1}\right)^{-b}\int_0^{u_1}du_2 \left(\frac{u_2}{1- u_2}
\right)^{b},
\end{equation}
which is equal to \cite{grad}
\begin{equation}
\label{J2b}
{\cal J}_2=\frac{- 36}{b^2(b+1)}\int_0^1 du_1 u_1
(1-u_1)^b\,{}_2F_1(b,b+1;b+2;u_1)\;.
\end{equation}
This last integral can also be done \cite{grad}, and we
finally obtain
\begin{equation}
\label{J2c}
{\cal J}_2=\frac{- 36}{b^2(b+1)^2(b+2)}\,\,
{}_3F_2(b,b+1,2;b+2,b+3;1)\;.
\end{equation}
%
%
%
%
%
After some rearrangement, we derive
\begin{equation}
\label{J2d}
{\cal J}_2=- 36\left[\zeta(2,b+1)+\frac{1}{2b^2}-\frac{1}{b}\right],
\end{equation}
involving Riemann's zeta function
$\zeta(z,q)\equiv \sum_0^\infty (n+q)^{-z}$. Plana's
formula of Appendix \ref{PLANA} may be used in the zeta
function, yielding
\begin{equation}
\label{ZETAPLANA}
\zeta(2,b+1)=\frac{1}{2(b+1)^2}+\frac{1}{(b+1)}+
\int_0^\infty \frac{dy}{e^{2\pi y}-1}\,
\frac{4(b+1)y}{[y^2+(b+1)^2]^2}\;,
\end{equation}
so that the result for ${\cal J}_2$ is
\begin{equation}
\label{J2e}
{\cal J}_2=- 36\left\{\frac{1}{2b^2(b+1)^2}
+\int_0^\infty \frac{dy}{e^{2\pi y}-1}\,\frac{4(b+1)y}
{[y^2+(b+1)^2]^2}\right\}.
\end{equation}
If we expand the last fraction in (\ref{J2e}) in a
power series in $y$, use the definition of the
Bernouilli numbers,
\begin{equation}
\label{BERNOUILLI}
\frac{B_{2n}}{4n}\equiv (-1)^{n-1} \int_0^\infty
\frac{dy\, y^{2n-1}}{e^{2\pi y}-1}\;,
\end{equation}
and the fact that $B_2=1/6$, we finally obtain (\ref{J2f}). 


\section{}
\label{EFFPOT}

We compute the effective potential at zero temperature for $d=3$, 
in order to illustrate the workings of the renormalization procedure. The
calculation can be viewed as a special case of the one performed in the text,
with the kink being traded for a constant background $\varphi_c$. In that
case, determinants are simple to compute, and we formally obtain:
\begin{equation}
\label{Veff}
V_{eff}(\varphi_c)=\frac{1}{4!}(\varphi_c^2-\varphi_v^2)^2 + \frac{\lambda}{2}
\int\frac{d^4 k}{(2\pi)^4}\ln\left[1+ \frac{\varphi_c^2-\varphi_v^2}
{2(k^2+M^2)}\right]\;.
\end{equation}
The subtraction of counterterms leads to:
\begin{equation}
\label{Veffct}
V_{eff}(\varphi_c)\to V_{eff}(\varphi_c) - \frac{1}{2}C_1(\varphi_c^2-\varphi_v^2)-
\frac{1}{4}C_2(\varphi_c^2-\varphi_v^2)^2\;,
\end{equation}
which will yield the renormalized expression if we use a cutoff $\Lambda$ to regularize (\ref{Veffct}), and fix renormalization conditions. Since we may write the renormalized
effective potential $V_R$ as:
\begin{equation}
\label{VRexp}
V_R(\varphi_c)=\sum_{n=1}^{\infty}\frac{1}{n!}{\widetilde\Gamma}^{(n)}_R
({\bf 0}, \cdots, {\bf 0})\, (\varphi_c-\varphi_v)^n\;,
\end{equation}
it is easy to show that the renormalization conditions (\ref{RN1}) and (\ref{RN2})
determine $C_1$ and $C_2$:
\begin{equation}
\label{C1}
C_1=\frac{\lambda}{2}\int^\Lambda \frac{d^4 k}{(2\pi)^4}\frac{1}{(k^2+M^2)}\;,
\end{equation}
\begin{equation}
\label{C2}
C_2=-\,\frac{\lambda}{4}\int^\Lambda \frac{d^4 k}{(2\pi)^4}\frac{1}{(k^2+M^2)^2}\;.
\end{equation}
Clearly, both expressions correspond to one-loop graphs at zero external momenta.
The renormalization procedure adopted in the text also uses zero-momentum 
subtractions in expansions in $(\varphi^2-\varphi^2_v)/2$,
but in the kink sector. In that case, the effective action in the kink 
sector replaces the effective potential, while the expansion (\ref{VRexp}) becomes a 
functional one 
\begin{equation}
\label{ARexp}
{\cal A}_R[\fith]=\sum_{n=1}^{\infty}\frac{1}{n!}\int \frac{d^4 p_1}{(2\pi)^4}\cdots
\int \frac{d^4 p_n}{(2\pi)^4} {\widetilde\Gamma}^{(n)}_R
({\bf p_1}, \cdots, {\bf p_n})\, [\fith({\bf p_1})-\varphi_v]\cdots[\fith({\bf p_n})-\varphi_v]\;,
\end{equation}
with $\fith({\bf p})$ standing for the Fourier transform of the kink background.


\end{document}